\documentclass[prl,aps,epsf,twocolumn,showpacs,10pt]{revtex4-2}
\usepackage{graphicx,amsfonts,times,bm,amsmath,verbatim,color,array}
\usepackage{natbib}

\begin{document}

\title{Nonreciprocal optics and magnetotransport in Weyl metals as signatures of band topology}

\author{S. Nandy}
\affiliation{Department of Physics, University of Virginia, Charlottesville, VA 22904, USA}
\author{D. A. Pesin}
\affiliation{Department of Physics, University of Virginia, Charlottesville, VA 22904, USA}


\begin{abstract}
We consider effects of spatial dispersion in noncentrosymmetric time-reversal invariant Weyl metals in the presence of a static magnetic field. In particular, we study currents that are linear in both the spatial derivatives of an applied electric field, and the static magnetic field, which are responsible for the phenomenon of gyrotropic birefringence. We show that the chiral anomaly and the chiral magnetic effect make the leading contribution to this class of phenomena in metals. We apply the obtained results to the problem of electromagnetic wave transmission through a thin slab of a Weyl semimetal, and show that the transmission coefficient contains a component that is odd in the applied static magnetic field. As such, it can be easily distinguished from conventional Ohmic magnetotransport effects, which are quadratic in the applied magnetic field. The relative magnitude of the effect can reach a few percent in Weyl materials subject to magnetic fields of 0.1Tesla, while the effect is several orders of magnitude smaller in metals without Berry monopoles. We conclude that the nonreciprocal optical and magnetotransport effects can be a robust probe of band topology in metals.
\end{abstract}

\maketitle


{\color{blue}{\em Introduction}}--- The current response to an electromagnetic plane wave in the linear response regime can be written through the conductivity tensor $\sigma_{\alpha \beta}({\mathbf{q},\omega})$ as
\begin{eqnarray}
j_{a}({\bm q,\omega})&&=\sigma_{ab}({\bm q,\omega})E_{b}({\bm q,\omega}),
\label{current_gen}
\end{eqnarray}
where $\omega$ and $\bm{q}$ are the frequency and the wave vector of the wave, respectively.

In the presence of a magnetic field, the conductivity tensor satisfies the Onsager symmetry relations,
\begin{align}
  \sigma_{ab}(\bm q,\omega,\bm B)=\sigma_{ba}(\omega,-\bm q,-\bm B).
\end{align}
These imply that in the expansion of the conductivity to linear order either in $\bm q$, or $\bm B$,
\begin{align}\label{eq:conductivityexpansion}
  \sigma_{ab}(\bm q,\omega,\bm B)\approx &\sigma^D_{ab}(\bm q,\omega)+\chi_{abc}(\omega)q_c+\lambda_{abc}(\omega)B_c\nonumber\\
  &+g_{abcd}(\omega)q_cB_d,
\end{align}
tensors $\sigma^D_{ab}$ and $g_{abcd}$ are symmetric in the first pair of indices, while tensors $\chi_{abc}$ and $\lambda_{abc}$ are antisymmetric in the first pair of indices. Of these, $\sigma^D_{ab}$ is the Drude conductivity, $\chi_{abc}$ and $\lambda_{abc}$ describe the natural optical activity, and the Hall effect. These two tensors are known to be sensitive to the band geometry. In particular, $\chi_{abc}$ and the associated phenomenon of the natural optical activity is related to the magnetic moments of quasiparticles~\cite{MaPesin2015,Zhong2016}, while $\lambda_{abc}$ and the associated Hall effect stem from the Berry curvature~\cite{NagaosaReview}.

Tensor $g_{abcd}$ is responsible for gyrotropic birefringence, and describes nonreciprocal propagation of waves in a material~\cite{Hornreich1968}. This tensor is the primary object of the present study. It is apparent that $g_{abcd}$ is an even-rank pseudotensor, therefore its existence requires broken inversion symmetry.

The nonreciprocal phenomena in insulating and molecular substances have a long history~\cite{Hornreich1968}, and have recently been re-formulated in the spirit of modern band theory~\cite{Malashevich}. Further developments pertaining to conducting systems~\cite{xiao2019nonreciprocal,Hughes2019semiclassical} showed that in such materials, the low-frequency nonreciprocal transport has geometric contributions related to the existence of an electric quadrupole moment of wave packets. Refs.~\cite{xiao2019nonreciprocal,Hughes2019semiclassical} focused on magnetic materials, and connected the aforementioned quadrupole moment to the dipole moment of the quantum metric tensor.

In this work, we show that in non-magnetic metals, but in the presence of an orbital magnetic field, there are the same type of nonreciprocal phenomena, which stem from the Berry curvature of the bands. We discuss how these phenomena manifest themselves in the transmission of electromagnetic waves through thin metallic films at low frequencies, and that they can serve as a robust test of band topology in metals. Physically, these effects can be thought of the electromagnetic analogs of the acoustic magnetochiral dichroism discussed in Refs.~\cite{garate2020phonon,antebi2020anomaly,sukhachov2020}, where it was shown that propagation of sound in a Weyl metal depends on whether it happens along or opposite to a magnetic field.

{\color{blue}{\em Nonreciprocal magnetotransport in Weyl metals}}--- In this work, our goal is to study how the effects of spatial dispersion associated with band topology affect optical properties of metallic thin films. Naturally, this requires studying propagation of electromagnetic waves in such films.  The effects of spatial dispersion are small in crystals due to the large value of the speed of light. For a wave with a phase velocity of $v_{\rm ph}$, one expects that the relative magnitude of the contribution of conduction electrons to spatial dispersion is $v_F/v_{\rm ph}$, where $v_F$ is the Fermi speed. Therefore, one should consider regimes of parameters in which the phase speed of the electromagnetic wave is as small as possible. In metals, this implies that one should consider the limit of low frequencies, in which electromagnetic waves penetrate the metal within the skin layer of width $\delta_s\sim 1/\sqrt{\mu_0\sigma\omega}$, and $v_F/v_{\rm ph}\sim v_F/\omega \delta_{s} \propto 1/\sqrt{\omega}$ for $\omega\ll1/\tau$, where $\tau$ is the transport mean free time.
%
%

At such low frequencies, we can consider only the intraband dynamics of electrons. In topological metals, the leading contribution to the nonreciprocal magnetotransport effects comes from the chiral anomaly and the chiral magnetic effect, which can be described using the macroscopic transport equation, as was previously discussed in Refs.~\cite{ParameswaranPesin2014,antebi2020anomaly}.

We note in passing that in metals without Weyl points, the gyrotropic birefringence effects are still possible, since their existence depends on the crystalline class of the material. However, the corresponding effects are much smaller than in Weyl materials at low frequencies. The difference is due to the fact that in regular noncentrosymmetric metals, even in the presence of Berry curvature, the current perturbations that it creates relax on the scale of the transport mean free time, while in Weyl materials the chiral anomaly leads to valley imbalances that relax on relatively long scale of intervalley scattering, which can be two orders of magnitude longer than the intravalley transport time. Therefore, in what follows we disregard the effects associated with the change in the valley density of states in a magnetic field~\cite{Xiao2010a}, as well as magnetic fields effect on impurity scattering. The latter should be important only in large magnetic fields, when Landau quantization is important.

We view a Weyl metal as a collection of Weyl nodes labeled with index $w$. Their chiralities $\eta_w$, are defined as the flux of the Berry  curvature, $\bm F_w$, through a closed surface in the momentum space, surrounding the corresponding node:
\begin{align}
\eta_w=-\frac{1}{2\pi} \oint d\bm S\cdot \bm {F}_w.
\end{align}
If the surface element $d\bm S$ is chosen to point along the outer normal for conduction bands, and along the inner normal for the valence bands, the chirality becomes a property of the node itself, and not of a particular band.

The Weyl nodes also have the Drude conductivity, $\sigma_{w,ab}$, and diffusion, $D_{w,ab}$, tensors. Here $a,b$ are Cartesian indices. The metal is subject to the electromagnetic field of a wave, the electric part of which we label $\bm E$, and an external static magnetic field $\bm B$. We will not explicitly refer to the magnetic field of the wave, so the notation for the external one should not lead to a confusion. The transport equation for the nonequilibrium density in valley $w$, $n_w$, and the Cartesian component $a$ of the electric current, $j_{w,a}$, in the same valley read
\begin{align}\label{eq:transporteq}
\partial_t n_w+\nabla_a j_{w,a}-\frac{e^2}{4\pi^2}\eta_w B_a E_a =-\frac{n_w}{\tau_v},
\end{align}
\begin{align}\label{eq:totalcurrent}
j_{w,a}=\sigma_{w,ab}E_b-eD_{w,ab}\nabla_b n_w+\frac{e}{4\pi^2}\eta_w B_a \frac{n_w}{\nu_w}.
\end{align}
The equations are standard by now, see \textit{e.g.} Refs.~\cite{ParameswaranPesin2014,antebi2020anomaly}. It is worth reminding that in Eq.~\eqref{eq:transporteq} for the density perturbation $n_w$, the third term on the left hand side represents the chiral anomaly, and the right hand side describes the phenomenological valley relaxation with $\tau_v$ as intervalley relaxation time. In Eq.~\eqref{eq:totalcurrent} for the electric current, the first two terms on the right hand side are the usual electric and diffusive currents, and the third term describes the chiral magnetic effect with $\nu_w$ as the density of states. We will discuss the inclusion of electronic screening below.

In what follows, we assume harmonic variation of $\bm E$ in space and time, $\bm E\propto e^{i\bm q\cdot\bm r-i\omega t}$. We solve Eqs.~\eqref{eq:transporteq} and~\eqref{eq:totalcurrent} to linear order in $\bm B$ and $\bm q$, and use the Einstein relation for the conductivity and diffusion tensors, $\sigma_{w,ab}=e^2\nu_w D_{w,ab}$, as well as the fact that the above two tensors are symmetric. By summing the current contributions from different nodes, we obtain the total electric current, $j_a=\sum_w j_{w,a}$, and the following expression for the tensor $g_{abcd}$ that determines the gyrotropic birefringence at low frequencies, see Eq.~\eqref{eq:conductivityexpansion}:
\begin{align}\label{eq:gtensor}
g_{abcd}=\frac{N_v e^3}{4\pi^2(\omega+\frac{i}{\tau_v})}\left(\tilde D_{ac}\delta_{bd}+\tilde  D_{bc}\delta_{ad}\right).
\end{align}
where $N_v$ is the number of valleys. To write $g_{abcd}$, we introduced the ``chiral diffusion tensor" $\tilde D_{ab}$, given by
\begin{align}\label{eq:chiralD}
\tilde D_{ab}=\frac1{N_v}\sum_w \eta_w D_{w,ab}.
\end{align}

Since we have limited ourselves to the low frequency regime, we need to take into account the electronic screening. At frequencies low compared to the plasma one, and at wave vectors small compared to the inverse Thomas-Fermi screening length, net density perturbations in a metal are forbidden, which is known as the electroneutrality limit. We can incorporate the effects of screening by subtracting the part of the nonlocal current that has nonzero divergence, as well as removing the longitudinal part of the external electric field, which is cancelled by the screening field. In other words, we make a simple redefinition
\begin{align}\label{eq:screenedg}
g_{abcd}\to \left(\delta_{aa'}-\frac{q_aq_{a'}}{q^2}\right)g_{a'b'cd}\left(\delta_{bb'}-\frac{q_bq_{b'}}{q^2}\right).
\end{align}
The projector on the left subtracts the part of the nonlocal current flowing along the direction of $\bm q$, while the one on the right subtracts the screened out longitudinal component of the external field. The resulting tensor $g_{abcd}$ remains symmetric with respect to the first pair of indices. In the specific calculation of wave transmission that we will undertake below, this redefinition is not important. However, we note in passing that if one assumes isotropic nodes, $\tilde D_{ab}\propto \delta_{ab}$, the resultant nonlocal current vanishes identically for transverse electromagnetic waves due to screening, see Eqs.~\eqref{eq:conductivityexpansion},~\eqref{eq:gtensor}, and~\eqref{eq:screenedg}.

Equations~\eqref{eq:gtensor},~\eqref{eq:chiralD} and~\eqref{eq:screenedg} are three of the main results of this work. The physical picture behind the two contributions to the right hand side of Eq.~\eqref{eq:gtensor} is as follows. The first one comes from the diffusive currents driven by nonuniform density perturbations in Weyl nodes due to the chiral anomaly. The second one comes from the chiral magnetic current driven by intravalley density perturbations caused by nonuniform electric currents driven by a nonuniform electric field. As expected from the Onsager relations, the resultant tensor $g_{abcd}$ is symmetric with respect to its first pair of indices.

\begin{figure}
\includegraphics[width=2.5in]{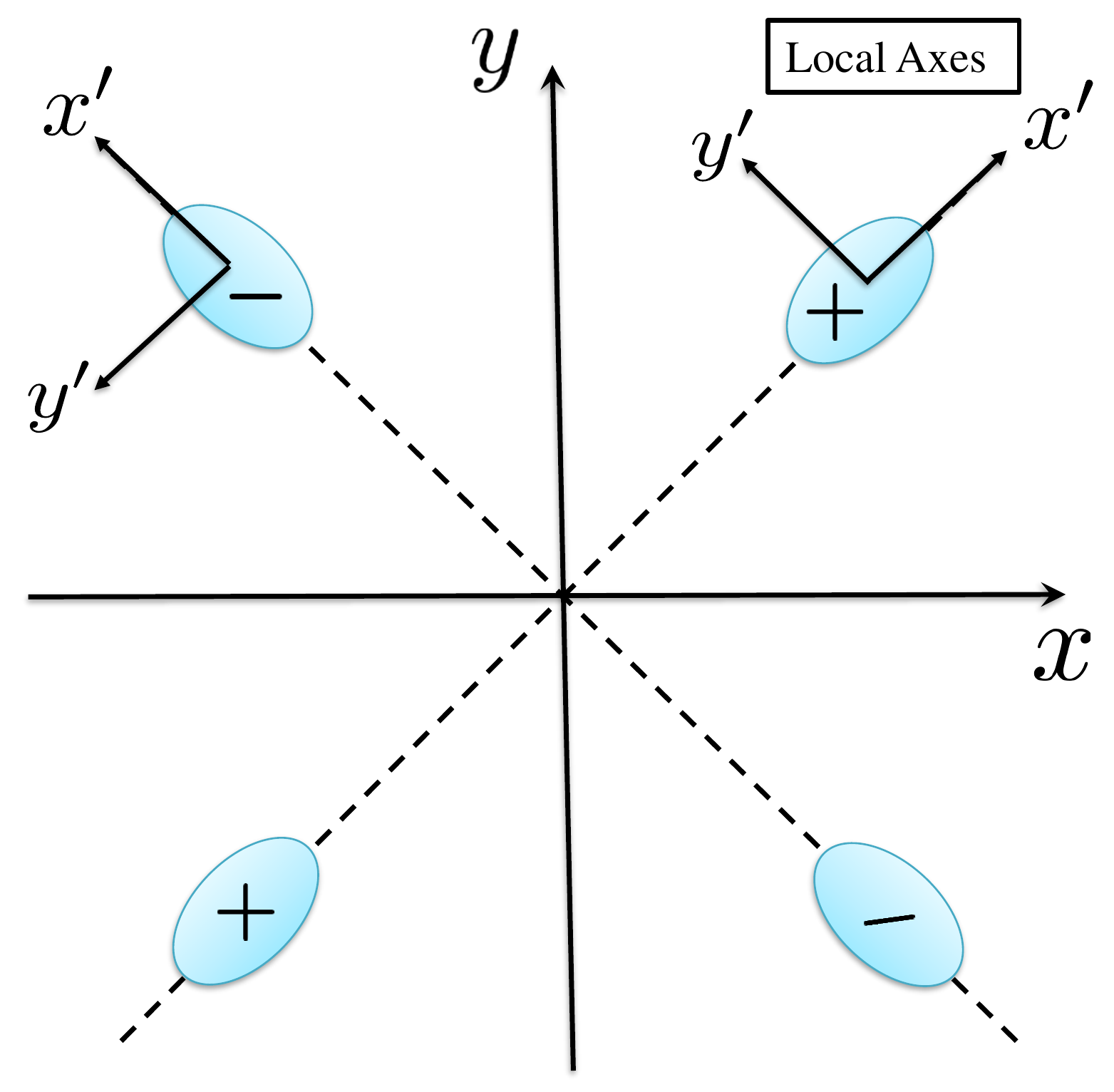}
\caption{The minimal model of a Weyl metal of $C_{2v}$ crystallographic class. The Weyl nodes of opposite chiralities (shown with ``+" and ``-" signs) are connected by mirror reflections in the $xz$ and $yz$ planes. Nodes of the same chirality are connected by the time reversal symmetry. The polar C$_{2}$ axis is perpendicular to the plane of the figure.}\label{fig:nodes}
\end{figure}

The band topology makes a contribution to the nonreciprocal magnetotransport through the chiral diffusion tensor, Eq.~\eqref{eq:chiralD}.  To illustrate how such an object appears, let us consider a hypothetical example of a Weyl metal with just four nodes, which belongs to the $C_{2v}$ crystallographic class. The Weyl nodes connected by the mirror operations of the $C_{2v}$ point group are shown in Fig.~\ref{fig:nodes}.  Being connected by the mirror reflections, the four nodes have essentially the same diffusion tensor, but written in different coordinate axes. The tensors of nodes of the same chirality, which are related by the time reversal symmetry, are identical. We assume that the diffusion tensors of individual valleys are diagonal in the coordinate axes aligned with the principal axes of the corresponding Fermi surfaces. Then in the global Cartesian coordinates we can write the following expressions for the two chiralities $\eta_{w}=\pm 1$:
\begin{align}
D_\pm=\left(
\begin{array}{ccc}
\frac{D_{x'x'}+D_{y'y'}}{2}&\pm\frac{D_{x'x'}-D_{y'y'}}{2}&0\\
\pm\frac{D_{x'x'}-D_{y'y'}}{2}&\frac{D_{x'x'}+D_{y'y'}}{2}&0\\
0&0&D_{zz}
\end{array}
\right),
\end{align}
which implies that the chiral diffusion tensor is
\begin{align}
\tilde D=\frac12(D_+-D_-)=\frac12(D_{x'x'}-D_{y'y'})\left(
\begin{array}{ccc}
0&1&0\\
1&0&0\\
0&0&0
\end{array}
\right).
\end{align}
This result is consistent with the general form of a second-rank pseudotensor in the $C_{2v}$ class~\cite{Malgrange2014}.

{\color{blue}{\em Propagation of electromagnetic waves through a thin topological metal film}}---The problem of electromagnetic wave propagation through a crystal of low symmetry can be made arbitrarily cumbersome. The combination of the crystal optics, scattering geometry, and a large variety of physical effects can make it very tedious to derive specific results. Below we aim to consider the simplest possible scattering problem of a linearly polarized wave transmission through a thin slab of material.

We consider the transmission of low-frequency transverse radiation perpendicular to the polar axis of a Weyl metal of $C_{4v}$ crystallographic class. This pertains to the monopnictide family of noncentrosymmetric Weyl metals~\cite{YanFelser2017review}. The sample is assumed to be under the conditions of the normal skin effect, such that the mean free path, $\ell$, and the skin depth, $\delta_s$ satisfy $\delta_s\gg \ell$. As in the rest of the paper, we assume that we can neglect diffusive relaxation of density perturbations. Physically, this means that the diffusive relaxation rate across the normal-skin depth is small compared to the wave frequency, or the intervalley scattering rate: $D\delta_s^{-2}\ll\mathrm{max}(\omega,1/\tau_v)$, where $D$ is the typical magnitude of the diffusion coefficient in a given valley. This condition is not restrictive for realistic impure samples. The problem of transmission through a slab of Weyl metal in an external magnetic field with arbitrary relationship between $D\delta_s^{-2}$ and $\omega$ was considered recently in Ref.~\cite{sukhachov2021transmission}. However, in that work, the nodes of the Weyl metal were assumed to be isotropic. As explained below Eq.~\eqref{eq:screenedg}, gyrotropic birefringence effects considered below cannot be obtained in such a model. Furthermore, in the Maxwell equations, we make the standard assumption that one can neglect the displacement current~\cite{LL8}. We also neglect the contribution of the surface states to electric currents flowing in the slab. Their existence does not lead to any nonreciprocity in transmission through the sample, while the effect on the overall transmission is small for macroscopic slabs.

\begin{figure}
\includegraphics[width=3in]{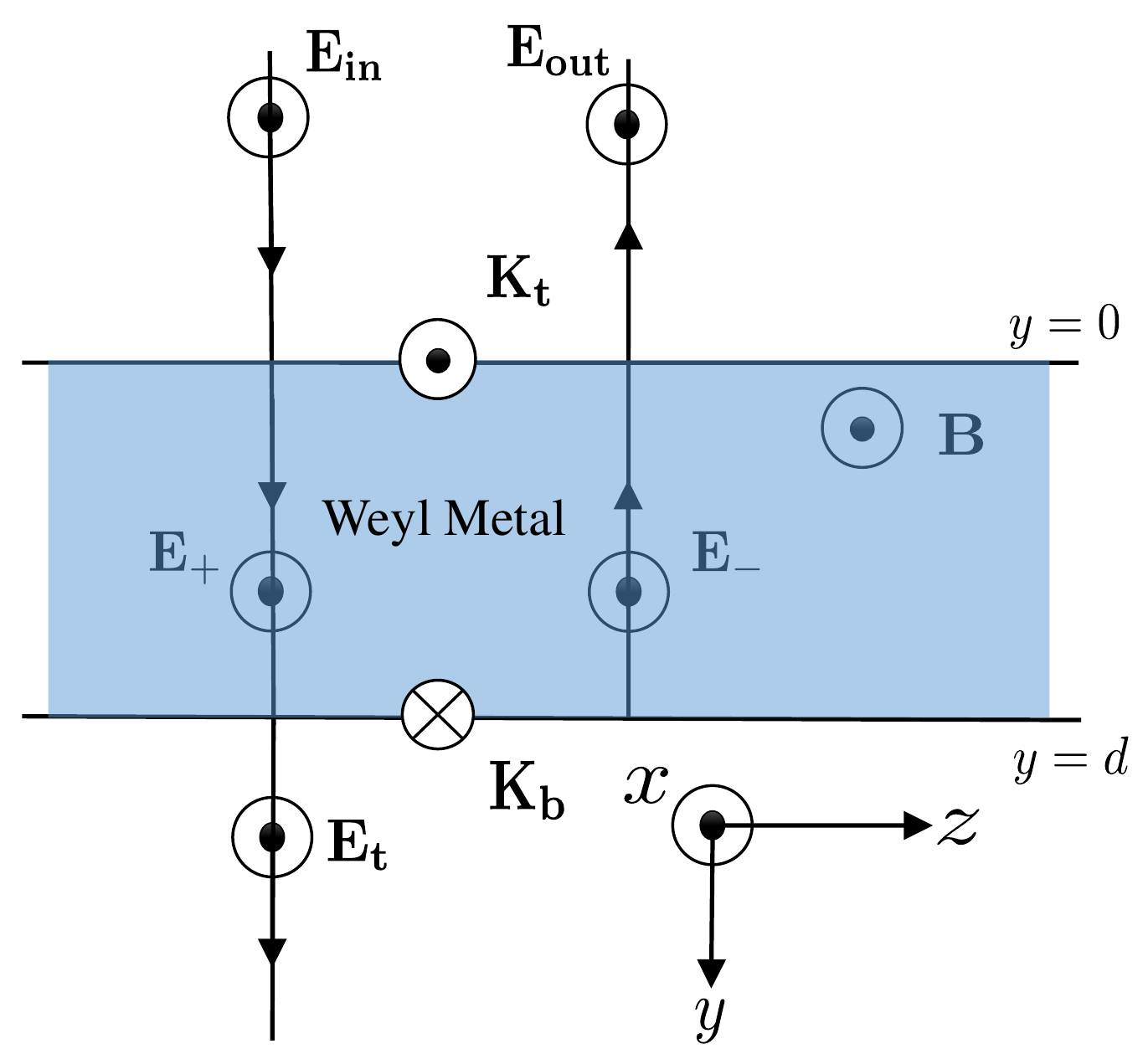}
\caption{Transmission of an electromagnetic wave though a slab of a Weyl metal of crystallographic class $C_{4v}$. The polar $C_4$ axis is directed along the $z$-axis. All waves inside and outside the slab are polarized in the $x$-direction. The external magnetic field $\bm B$ is also in the $x$-direction. The effective surface currents on the top and bottom surfaces, $\bm K_{t,b}$, flow in the opposite directions.}\label{fig:transmission}
\end{figure}
For the $C_{4v}$ crystallographic class, the chiral diffusion tensor $\tilde D_{ab}$ in Eq.~\eqref{eq:gtensor} has a single independent component, $\tilde D_{xy}=-\tilde D_{yx}\equiv \tilde D$. If we define
\begin{align}
  g=\frac{N_v e^3\tilde D}{4\pi^2(\omega+\frac{i}{\tau_v})},
\end{align}
we can write the total current in the bulk of the sample for a wave with a wave vector $\bm q=(0,q_y,0)$ and polarization vector $\bm E$ as follows:
\begin{align}\label{eq:bulkcurrent}
\bm j^{bulk}=\sigma^D \bm E+g q_y (\bm E\cdot \bm B) \bm e_x+g q_y E_x \bm B.
\end{align}
In this expression, we did not include the contribution of the ``dynamic chiral magnetic effect"~\cite{MaPesin2015,Zhong2016}, or simply the natural optical activity part of the nonlocal current, which is represented by tensor $\chi_{abc}$ in Eq.~\eqref{eq:conductivityexpansion}. Our goal is to describe the part of the transmission probability that is nonreciprocal, or, equivalently, odd in the external magnetic field $\bm B$. Natural optical activity has got nothing to do with such transmission. We also did not include the Hall effect, described by tensor $\lambda_{abc}$ in Eq.~\eqref{eq:conductivityexpansion}, which simply vanishes for the geometry we are going to consider.

In the presence of boundaries, it is necessary to consider $g=g(y)$ for the geometry of Fig.~\ref{fig:transmission}, since $g$ has to vanish outside the sample. In this case, one can no longer use Fourier representation for the current density. It is also~\textit{a priori} unclear in which order one has to apply $g(y)$ and spatial gradients to the electric field to obtain the current. It was shown in Ref.~\cite{agranovichyudson} that the prescription consistent with the Onsager reciprocity relations, and appropriate for the present geometry, is $gq_y\to \frac{1}{2i}(\partial_y g(y)+g(y)\partial_y)$. In the first term on the right hand side, the derivative acts on the product of $g(y)$ and the electric field, which is not written explicitly on the right. This recipe implies the existence of a surface current term, which is proportional to $E_x \partial_y g(y)$, which alters the boundary conditions for the magnetic field~\cite{agranovichyudson,HosurKerr2014}, see below.

In what follows we specialize to a B-field oriented along the $x$-axis, $\bm B=B(1,0,0)$. Then a wave polarized along the $z$-axis is not affected by nonreciprocal effects, $\bm j=\sigma^D \bm E$ for this polarization, see Eq.~\eqref{eq:bulkcurrent}. We wrote the Drude conductivity tensor without indices, but $\sigma^D_{xx}$ is implied. For a wave polarized along the $x$-axis, we obtain
the total current in the following form:
\begin{align}\label{eq:jx}
j_x=\sigma^D E_x-2igB \partial_y E_x -igBE_x\left(\delta(y)-\delta(y-d)\right).
\end{align}
The two last terms, which contain the spatial $\delta$-functions, represent the effective surface currents on the top and bottom surfaces of the sample, $\bm K_{t,b}$, respectively:
\begin{align}\label{eq:surfacecurrent}
  \bm K_{t}=-igBE_x(y=0)\bm e_x, \,\, \bm K_{b}=igBE_x(y=d)\bm e_x.
\end{align}
The existence of these surface currents implies that the component of the \emph{macroscopic} magnetic field along the surface is not continuous. There is no ambiguity in the boundary condition for the magnetic field, since the electric field is continuous across the boundary, so the coefficient in front of the $\delta$-functions is the same inside and outside the sample (by construction, $g$ there is the value inside the sample). It is also the surface currents of Eq.~\eqref{eq:surfacecurrent} that ensure that the reflection coefficient for normal incidence is an even function of the external magnetic field~\cite{Halperin1992reciprocity,Shelankov1992reciprocity}.

The rest of the considerations proceed in the standard way~\cite{LL8}. We use the bulk part of the current, Eq.~\eqref{eq:bulkcurrent}, or the first two terms on the right hand side of Eq.~\eqref{eq:jx}, to determine the bulk modes in the slab. We denote these modes with subscripts $\pm$ for propagation directions along and opposite to the $+y$-direction, see Fig.~\ref{fig:transmission}.  For a given frequency $\omega$, they have the form of plane waves, $\bm E_{\pm} e^{i q_{\pm}(\omega) y}$, with magnetic field amplitudes determined from the Faraday's law. The wave numbers $q_{\pm}(\omega)$ depend on the wave propagation direction, and to linear order in $\bm B$ are given by
\begin{align}\label{eq:qy}
  q_{\pm}(\omega)=i \mu_0gB\omega\pm (1+i)\frac{1}{\delta_s},
\end{align}
where $\delta_s=\sqrt{2/\mu_0\sigma^D\omega}$ is the skin depth. In the right hand side of Eq.~\eqref{eq:qy}, the ``+'' sign is chosen for modes propagating along the positive $y$-direction, and the ``-" sign is chosen for the negative $y$-direction to make sure that the waves attenuate as they propagate. If we separate the real and imaginary parts of $g$, such that $g=g'+i g^{''}$ with real $g^{'}$ and $g^{''}$, it is clear from Eq.~\eqref{eq:qy} that $g'$ determines the nonreciprocal attenuation, that is, the difference between the attenuation coefficients for the two opposite propagation directions, while $g^{''}$ would have described the difference in the refractive indices, were the wave not overdamped due to the Ohmic losses.

Having determined the bulk modes, we match the components of the electric and magnetic fields along the surface of the slab, taking into account surface currents of Eq.~\eqref{eq:surfacecurrent}. Omitting the algebra, we present the result for the odd part of the transmission coefficient, $T(\bm B)=|E_t(\bm B)/E_{in}|^2$, as a function of the external magnetic field:
\begin{align}
  \left|\frac{T(\bm B)-T(-\bm B)}{2T(0)}\right|=\sinh\left(2\mu_0g'B\omega d\right).
\end{align}
This is another main result of this work, which relates the topology of the band structure, encoded in $g'$, to a physical property: the linear in the external magnetic field part of the transmission coefficient. The dimensionless quantity that determines this transmission asymmetry is given by
\begin{align}\label{eq:estimate}
  2\mu_0g'B\omega d=\frac{\mu_0 N_v e^3\tilde D Bd}{2\pi^2}\frac{\omega^2\tau_v^2}{(1+\omega^2\tau_v^2)}
\end{align}
We can estimate its magnitude if we make a realistic assumption that the anisotropy of the diffusion tensor is strong, $\tilde D\sim v_F^2\tau$. Restoring the Planck's constant for a moment, and setting $1/\tau_v\lesssim\omega\lesssim1/\tau$, we obtain
\begin{align}
  \mu_0g'B\omega d\sim \frac{e^2}{4\pi\epsilon_0\hbar c}\frac{v_F}{c}\frac{B \ell d}{\Phi_0}N_v.
\end{align}
In the above expression, $c$ is the speed of light in vacuum, $\ell=v_F\tau$ is a length scale that roughly corresponds to the transport mean free path, and $\Phi_0=h/e$ is the flux quantum. The first two factors are small, their product is of order of $10^{-5}$, while the third one can be very large even for classically moderately strong fields. The numerical value of this parameter can be estimated by using typical values of parameters for TaAs with $N_v\sim 20$: $v_F\sim 3\times10^5\mathrm{m/s}$, $\tau\sim 10^{-12}\mathrm{s}$, such that $\ell\sim 0.3\mathrm{\mu m}$. Then for $\omega\sim 10^{11}\mathrm{rad/s}$ and $\sigma^D\sim 10^5\mathrm{\Omega^{-1}m^{-1}}$ we obtain $\delta_s\sim 10\mathrm{\mu m}$. We note that $\delta_s/\ell\sim 30$, and the diffusive limit that we assumed is well satisfied. Then for $B=0.1\mathrm{T}$, and $d\sim \delta_s$ we get $B \ell d/\Phi_0\sim 10^2$. This sets the relative magnitude of nonreciprocal phenomena at $1\%$. This is a very large value, as compared to gyrotropic birefringence effects in transparent magnetic insulators, which are weaker by at least two orders of magnitude ~\cite{pisarev1994optics}.

{\color{blue}{\em Conclusions}}--- In this work we demonstrated that the chiral anomaly and the chiral magnetic effect lead to gyrotropic birefringence and nonreciprocal propagation of electromagnetic waves in anisotropic topological metals. The effect comes from nonuniform in space redistribution of carriers among valleys due to the chiral anomaly driven by a nonuniform  electric field, and due to the chiral magnetic effect driven by valley-dependent charge accumulation. Similar mechanisms of current generation in a magnetic field were considered in Ref.~\cite{sukhachov2021transmission}, where it was shown that in the isotropic model of a Weyl metal there are nonlocal electric currents quadratic in the external magnetic field. In contrast, it is shown in this work that in anisotropic Weyl metals the chiral anomaly and the chiral magnetic effect lead to nonlocal electric currents \emph{linear} in the external magnetic fields.
 
Linear in the external magnetic field nonreciprocal transport and optical effects can be a robust signature of band topology in metals. Their magnitude in Weyl metals, albeit not large in absolute terms -- about 1\% for $B\sim0.1\mathrm{T}$ --  is nevertheless much greater than that of their counterparts in conventional materials~\cite{pisarev1994optics}. They are also free from the usual problem of separating the transport features due to the chiral anomaly that are quadratic in the magnetic field~\cite{SonSpivak} from mundane Ohmic effects~\cite{hassinger2016jetting,ong2021anomaly}. In this regard, the present work is similar in spirit to the proposals of Refs.~\cite{ParameswaranPesin2014} and~\cite{nandypesin2020}, but is free from difficulties associated with multi-terminal transport geometries, or strong-electric-field transport.  Using nonreciprocal optical and transport effects as a test of band topology in metals can be hindered by their possible polycrystallinity, which randomizes the orientation of the optical axes, and removes the nonreciprocal effects in bulk samples. Practically, this limits the value of the slab width in Eq.~\eqref{eq:estimate}.

This work was supported by the National Science Foundation Grant No. DMR-2138008.

\bibliography{nonreciprocal_references}
\bibliographystyle{apsrev}

\end{document}